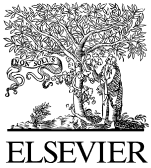

# The physics of solar spectral imaging observations in dm-cm wavelengths and the application on space weather

Baolin Tan [a,b,c,*], Yihua Yan [b,c], Jing Huang [a,b,c], Yin Zhang [a,b], Chengming Tan [b,c], Xiaoshuai Zhu [b]

[a] *National Astronomical Observatories of Chinese Academy of Sciences, Datun Road A20, Chaoyang District, Beijing 100101, China*
[b] *State Key Laboratory of Solar Activity and Space Weather, National Space Science Center of Chinese Academy of Sciences, Beijing 100084, China*
[c] *School of Astronomy and Space Sciences, University of Chinese Academy of Sciences, Beijing 100049, China*

**Abstract**

Recently, several new solar radio telescopes have been put into operation and provided spectral-imaging observations with much higher resolutions in decimeter (dm) and centimeter (cm) wavelengths. These telescopes include the Mingantu Spectral Radioheliograph (MUSER, at frequencies of 0.4–15 GHz), the Expanded Owens Valley Solar Array (EOVSA, at frequencies of 1–18 GHz), and the Siberian Radio Heliograph (SRH, at frequencies of 3–24 GHz). These observations offer unprecedented opportunities to study solar physics and space weather, especially to diagnose the coronal magnetic fields, reveal the basic nature of solar eruptions and the related non-thermal energy release, particle accelerations and propagation, and the related emission mechanisms. These results might be the important input to the space weather modeling for predicting the occurrence of disastrous powerful space weather events. In order to provide meaningful reference for other solar physicists and space weather researchers, this paper mainly focus on discussing the potential scientific problems of solar radio spectral-imaging observations in dm-cm wavelengths and its possible applications in the field of space weather. These results will provide a helpful reference for colleagues to make full use of the latest and future observation data obtained from the above solar radio telescopes.
© 2022 COSPAR. Published by Elsevier B.V. All rights reserved.

## 1. Introduction

Solar radio astronomy is a burgeoning branch of astrophysics affecting almost all aspects of solar physics, which include the physics of the quiet Sun and various kinds of solar activity in the huge space from the photosphere, through the chromosphere to the high corona, and even far into interplanetary space. The related emission frequency ranges from several hundreds of GHz (sub-millimeter wavelength) down to sub-MHz (kilometer wavelength) (Dulk, 1985; Bastian et al., 1998). As we know, the emission at different frequencies generally comes from different levels of the solar atmosphere. For example, the emission at sub-millimeter (sub-mm) and millimeter (mm) wavelengths may be generated near the photosphere and lower chromosphere, decimeter (dm) and centimeter (cm) wavelengths may be produced in the higher chromosphere and corona, meter wavelengths may come from the higher corona, and decameter and hectometer wavelengths are usually associated with interplanetary space. Because the most remarkable phenomena in solar physics, such as the origin and triggering of various kinds of solar eruptions (including solar flares, coronal mass ejections, eruptive filaments, jets, etc.), coronal heating, and even the source region of solar wind take place almost exclusively in the chromosphere or lower corona, dm-cm wavelength (from several hundred MHz to beyond 10 GHz) emissions are most sensitive to the above solar phenomena. Therefore, the high performance solar radio observations in dm-cm wavelengths have become particularly important which has the potential to reveal the rhythms of solar eruptions, unlock the mystery of corona heating, and provide more important information for modeling and predicting the occurrence of the disastrous space weather events.

* Corresponding author at: National Astronomical Observatories of Chinese Academy of Sciences, Datun Road A20, Chaoyang District, Beijing 100101, China.
*E-mail address:* bltan@nao.cas.cn (B. Tan).

The solar radio observations in dm-cm wavelengths can be obtained simply from single-dish radio telescopes which may provide solar total emission intensity at single frequency or dynamic spectrum in a broadband frequency range. From the observations of broadband dynamic spectrometers, for example, the Ondrejov Radiospectrograph in the Czech Republic (ORSC, Jiricka et al., 1993), the Chinese Solar Broadband Radio Spectrometers (SBRS, Fu et al., 2004), etc., many new properties of solar radio bursts with superfine structures are reported, such as the distribution of separation frequency of microwave type III pairs (Aschwanden and Benz, 1997; Tan et al., 2016a), statistical characteristics of Zebra patterns (Chernov, 2010; Tan et al., 2014) and fiber bursts (Karlicky et al., 2013; Wan et al., 2021), evolutionary characteristics of microwave spike groups in the flaring processes (Wang et al., 2008; Chernov et al., 2010; Tan, 2013; Tang et al., 2021), multi-timescale microwave quasi-periodic pulsations (QPPs, Tan et al., 2010), and various bizarre spectral fine structures occurred around flare events (Chernov, 2011). However, without spatial resolution, we don't know exactly where the source region of the solar radio emission or bursts are, what is the relationship between the radio bursts and the magnetic field configuration, and what is the real formation mechanism of various solar radio bursts with complex and superfine structures (Chernov et al., 2014).

Recently, several new solar radio telescopes have been put into operation and provided solar spectral-imaging observations with high resolutions in the range of dm-cm wavelengths. Such telescopes include,

(1) Mingantu Spectral Radioheliograph (MUSER). It is located in China and the frequencies cover 0.4–15 GHz. The frequency resolution is 25 MHz. The temporal resolution (cadence) is 25 ms in the frequency range of 0.4–2.0 GHz and 0.2 s in the frequency range of 2.0–15.0 GHz. The spatial resolution is 1.3–63 arcsecond depending on the frequencies (Yan et al., 2021).

(2) Expanded Owens Valley Solar Array (EOVSA). It is located in America and the frequencies cover 1–18 GHz and the temporal resolution is about 1 s at more than 100 frequency channels (Gary et al., 2018).

(3) Siberian Radio Heliograph (SRH). It is located in Russia and the frequencies cover 4–8 GHz with superhigh sensitivity of about 0.01 sfu, spatial resolution of 1–2 arc minutes and imaging interval of about 12 s (Lesovoi et al., 2012; Grechnev et al., 2018). Recently, SRH is expanded to cover the frequency range 3–24 GHz and is still in its commissioning phase (Altyntsev et al., 2020).

The above new instruments provide an unprecedented opportunity to meet the above requirements of solar physics and space weather. This work attempts to present and discuss the main problems which are related to various solar activities and the possible application on space weather. The paper is organized as following: Section 2 discusses the main solar physical problems which may require observations of broadband spectral-images with high resolutions in dm-cm wavelengths. Section 3 presents the possible applications of the research and observations with radio broadband spectral-images in dm-cm wavelengths. Finally, conclusions and discussions are summarized in Section 4.

## 2. The main solar physical problems faced by spectral imaging observations in dm-cm wavelengths

In the past several decades, with the improvement of the spectral and temporal resolutions of the solar radio telescopes, many unprecedented features of the fine structures on the broadband dynamic spectrum have been studied at dm-cm wavelength, such as fast QPPs, microwave Zebra patterns, fibers, spike burst groups, type III groups, and many bizarre spectral fine structures. From the spectrogram, we may extract many parameters, including the emission intensity, lifetime, frequency bandwidth, polarization degree, frequency drifting rate, etc (Tan, 2008). However, because these instruments have no spatial resolutions, we can not obtain the information about the source regions, including their position, scales, structures, and the relationships with certain magnetic field configurations. Therefore, so far, we do not fully understand the origin and formation mechanism of the various kinds of solar radio bursts. With the advent of MUSER, EOVSA and SRH, we will obtain a completely new understanding of the above solar radio bursts. This section will discuss the main solar physical problems from seven aspects that are currently attracting much attention and will be greatly advanced by the new radio spectral imaging observations. We hope this discussion can provide a meaningful reference for other solar physicists and space weather researchers.

(1) Diagnosing the magnetic fields in the chromosphere and corona.

The magnetic field plays a key role in various solar physical processes, including the origin and evolution of solar eruptions, the generation and maintenance of hot corona, and even the formation of the rich structures of the solar atmosphere. It can be directly measured by using the Zeeman effect at optical wavelengths in the photosphere and low chromosphere. However, in the high chromosphere and corona, we can not obtain the magnetic field by using the same way. Radio observations are considered to be the most important method to diagnose the magnetic fields in the high chromosphere and corona.

There are many people whose works have greatly advanced the radio diagnosis of magnetic fields in chromosphere and corona (Dulk, 1985; Bastian et al., 1998; Gelfreikh et al., 1987; Kundu, 1990; Zhou and Karlicky, 1994; Huang, 2006; Huang et al., 2013; Wiegelmann et al., 2014; Yan et al., 2020, etc.). There are several review papers that elaborate on this topic (Dulk and McLean, 1978; Kundu, 1990; Kruger and Hildebrandt, 1993; White, 2005; Casini et al., 2017; Alissandrakis and Gary, 2021). Many people have attempted to derive the chromospheric and coronal magnetic fields directly from the radio observations obtained by Nobeyama Radioheliograph or EOVSA (Huang, 2006; Huang et al., 2013; Zhu et al., 2021; Chen et al., 2020; Fleishman et al., 2020, etc.).

One of the recent results is shown in Fig. 1 which is obtained by Fleishman et al. (2020). Here, the coronal magnetic field maps are derived for the X8.2 flare on 10 September 2017 from the radio observation of EOVSA in 26 microwave bands in the frequency range of 3.4–15.9 GHz, which demonstrated the fast evolution of the flaring coronal magnetic field. They use a cost function combining the gyrosynchrotron (GS) emission and free-free bremsstrahlung emission and a Fast Gyrosynchrotron Codes (Fleishman and Kuznetsov, 2010) to derive the magnetic field around the flaring source region. They found that the decay of the coronal magnetic field could release sufficient energy to power a strong flare event.

The other result is shown in Fig.2 which is obtained by Chen et al. (2020). They obtained the magnetic field strength and the energetic electron density with energy above 300 keV at different heights along a flare current sheet. These results are derived from the microwave images by EOVSA at frequency of 3.4–15.9 GHz. Here, the magnetic field strengths are derived by using best-fitting based on gyrosynchrotron radiation. The measured magnetic field profile shows a reconnection X point where the reconnecting field lines of opposite polarities closely approach each other.

These successful results obtained by EVOSA once again show the importance of the broadband spectral radio imaging observations with high spatial–temporal-frequency resolutions.

Practically, because of coexistence of multiple emission mechanisms (including bremsstrahlung emission, cyclotron, gyro-synchrotron and synchrotron emissions, and even coherently plasma emission) in the high chromosphere and low corona from the quiet region via active region to flaring source region, and when different mechanisms are dominant, the dependence between the magnetic fields and observable radio parameters is also different. For example, even the region around a flare current sheet, the radio emission possibly not only has the contribution of gyrosynchrotron radiation, but also may partly come from coherent plasma emission. Therefore, we have to determine the dominant emission mechanism at each diagnosing pixel, and then adopt the appropriate diagnostic function to retrieve the magnetic fields. Recently, we collected and sorted out all methods for diagnosing coronal magnetic fields from solar radio observations, including some important methods neglected in previous works (for example, the difference of cutoff frequencies between left- and right-handed circular polarization emission can be applied to measure the magnetic field and plasma density, etc.), and formed a complete assembly of explicit radio diagnostic functions of coronal magnetic fields, which covers the bremsstrahlung emission of thermal plasma above the Sun's quiet region, cyclotron and gyro-synchrotron emissions from magnetized hot plasma and mildly relativistic nonthermal electrons above the active region, synchrotron emission from relativistic high energy electrons and coherently plasma emission around the flaring regions (Tan, 2022).

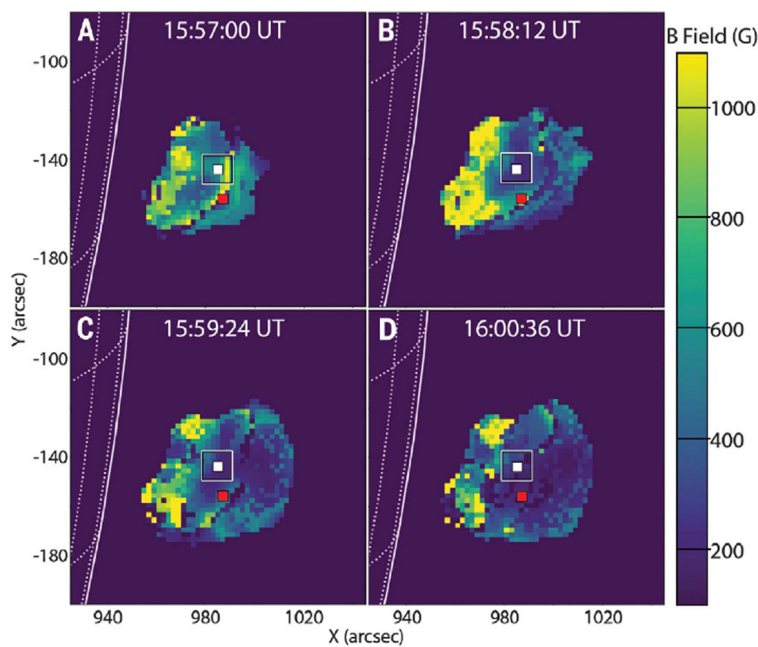

Fig. 1. The coronal magnetic field maps derived around the flaring source region of the X8.2 flare on 10 September 2017 from the radio observation obtained by the Expanded Owens Valley Solar Array (EOVSA) in 26 microwave bands in the range of 3.4–15.9 GHz. Courtesy of Fleishman et al. (2020).

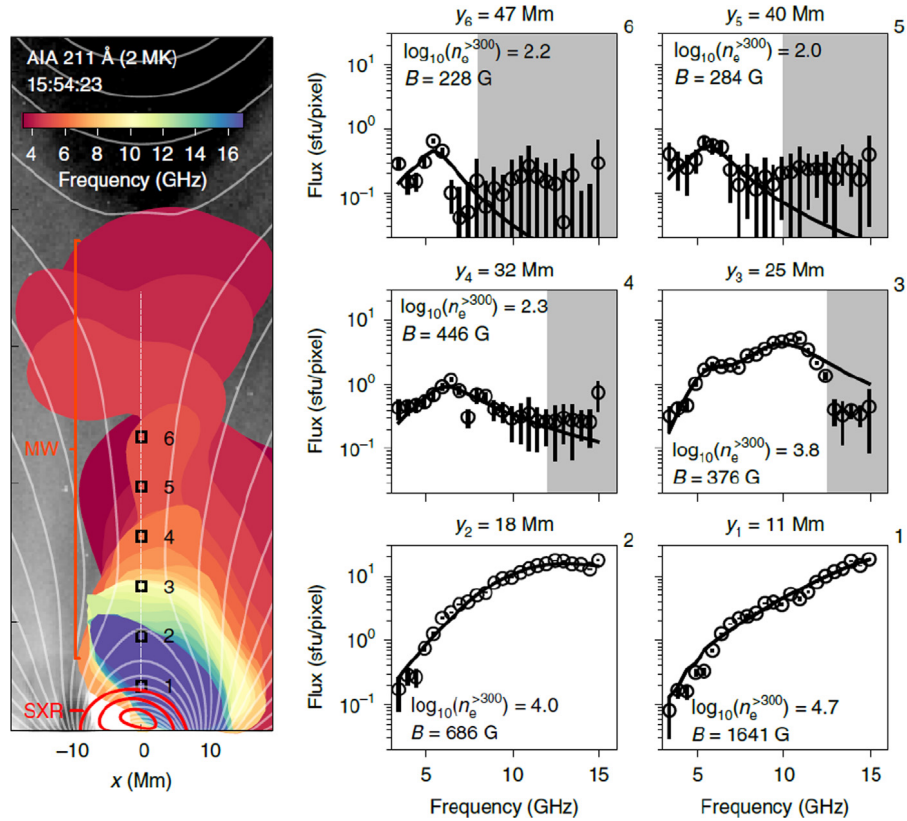

Fig. 2. The measurement of the spatially resolved magnetic field and flare-accelerated relativistic electrons along a current-sheet in a solar flare. The left panel shows the microwave images at frequency of 3.4–15.9 GHz overlapped on AIA EUV image at 211 Å. The right panel presents the magnetic field strength and energetic electron density at 6 different heights along the flare current sheet. Courtesy of Chen et al. (2020).

In fact, there are still many aspects of the above diagnosing functions that deserve modification and improvement in the future works, for example, the diagnostics of spectral fine structures (such as coherent emission associated to microwave Zebra patterns, fibers, spike bursts, etc.) in the flaring source regions depends heavily on the theoretical models. Some models are still immature which need further improvement. Obviously, different diagnostic functions are applicable to different physical conditions. On the basis of the above set of explicit radio diagnostic functions, combined with the new spectral imaging observations, it is possible to obtain magnetic maps of the solar full-disk at different heights in the chromosphere and corona with relatively high spatial resolutions. This will greatly promote the exploration of the major issues related to the origin of solar bursts and the mystery of coronal heating.

(2) The generation of microwave quasi-periodic pulsations (QPPs).

Solar quasi-periodic pulsations (QPPs) are frequently observed in multiple frequency band emissions (Aschwanden, 1987; Nakariakov and Melnikov, 2009). In the same way that seismic waves can probe the interior structures of the Earth, solar QPP can be used to diagnose structural and physical parameters of solar source regions. Especially the flare-associated microwave QPPs with pulsating periods range from tens of milliseconds to several minutes (Melnikov et al., 2005; Nindos and Aurass, 2007; Kupriyanova et al., 2010; Tan et al., 2010; Tan and Tan, 2012; Huang et al., 2014, etc.) can provide the important information of the microphysics of energy release, magnetic reconnection, and particle accelerations in flaring source regions. For example, Kashapova et al. (2021) analyzed the QPP detected in the microwave and dm emission of a flare on 5 May 2017 using simultaneous observations of SRH and MUSER, shown in Fig. 3. They found that the period of the microwave QPP is about 30 s and comes from typical gyrosynchrotron emission modulated by the standing-sausage oscillation and the related nonlinear acceleration processes in the current sheet of a compact loop-like source region with a typical height of about 31 Mm, while the simultaneous dm QPP exhibited a period of only 6 s and came from coherent plasma emission. However, as lack of imaging observation by MUSER at dm wavelengths in this event, we do not known where is the source region of dm QPP and what is its exact link with the above microwave QPP.

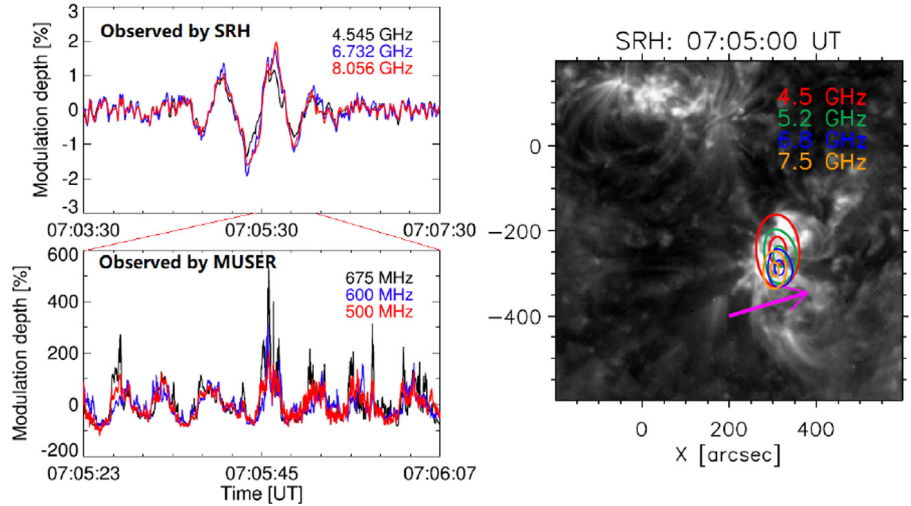

Fig. 3. The spectral and imaging observations of a QPP in a flare event by SRH and MUSER. The left panels show the de-trended oscillation temporal profiles at microwave (upper, observed by SRH) and dm bands (lower, observed by MUSER), the right panel presents the contours of QPP source region at different microwave frequencies (observed by SRH). courtesy to Kashapova et al. (2021).

Based on the multi-wavelength joint-imaging observations of AIA/SDO, MUSER, Nobeyama Radioheliograph and RHESSI, Chen et al. (2019a) found there are three QPP components that occurred with the M8.7 flare on 2014 December 17: 4-min UV QPP at 1600 Å images in the center of the active region from the preflare phase to the impulsive phase, 3-min EUV QPP along the circular ribbon during the preflare phase, and 2 min radio QPP that occurred around the two-ribbon flaring region during the impulsive phase. The source region of the radio QPP emission is shown in Fig. 4. Further analysis indicated that the 3-min EUV QPP is closely linked to the 2 min radio QPP and they provide possible implications for flare precursors.

Limited by the resolutions of the previous observations, so far, the formation of broadband radio QPPs is not quite clear. Especially the formation mechanisms of fast microwave QPPs with periods of sub-second (Tan et al., 2010, etc.) and the complex microwave QPP with zigzag pattern (Z-QPP) in the frequency range of 1.1–1.34 GHz shown in Fig. 5. The Z-QPP clearly consists of three sections: the first section (QPP1 in Fig. 5) starts at 06:04:32 UT and ends at 06:04:43 UT with global frequency drifting rate of $-16.5$ MHz/s, period of about 60 ms, and the bandwidth increasing slowly from 60 MHz to 160 MHz. The second section (QPP2 in Fig. 5) starts at 06:04:43 UT and ends at 06:04:47 UT without global frequency drifting rate, period of about 75 ms, and bandwidth of 140–200 MHz with few variations. The third section (QPP3 in Fig. 5) starts at 06:04:47 UT and ends at 06:04:57 UT with global frequency drifting rate of about $-5.6$ MHz/s, period of about 55 ms, and bandwidth increasing slowly from 60 to 140 MHz, and possibly beyond the observed frequency domain. We believe that they should carry important information about the dynamic processes in the source region which can be revealed from the new spectral imaging observations with high resolutions. This information includes the locations, spatial structure, motion and changes over time of the pulsating source regions. They may help us to understand the basic process of energy release, the triggering and evolution of the eruptions, particle acceleration and propagations.

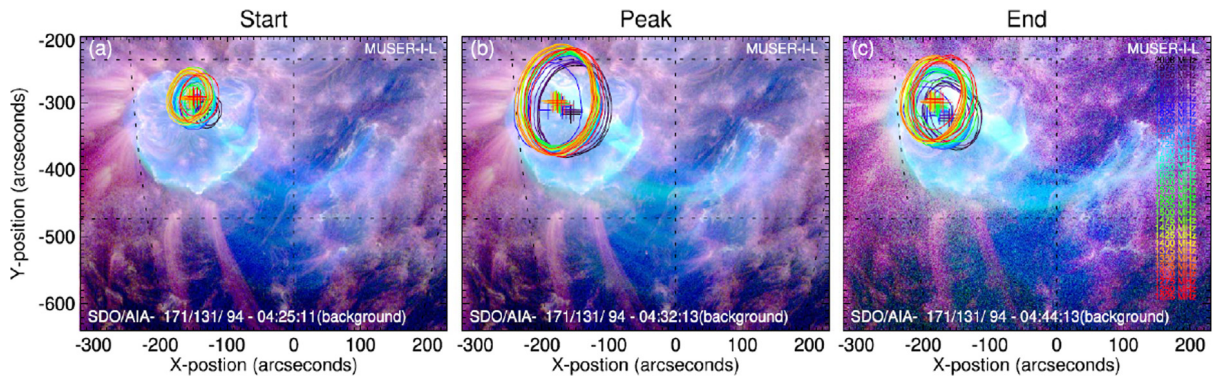

Fig. 4. Images of the radio source observed by MUSER at different frequencies from 1.2 to 2.0 GHz, with a frequency resolution of 25 MHz in a M8.7 flare on 2015 December 17. (a) - (c) Different color contours of the radio source with centroids (plus signs) at a fixed intensity, overlaid on the AIA images at the star (04:25 UT), peak (04:32 UT), and end (04:44 UT) times of the radio bursts associated to the radio QPP. Courtesy to Chen et al. (2019a).

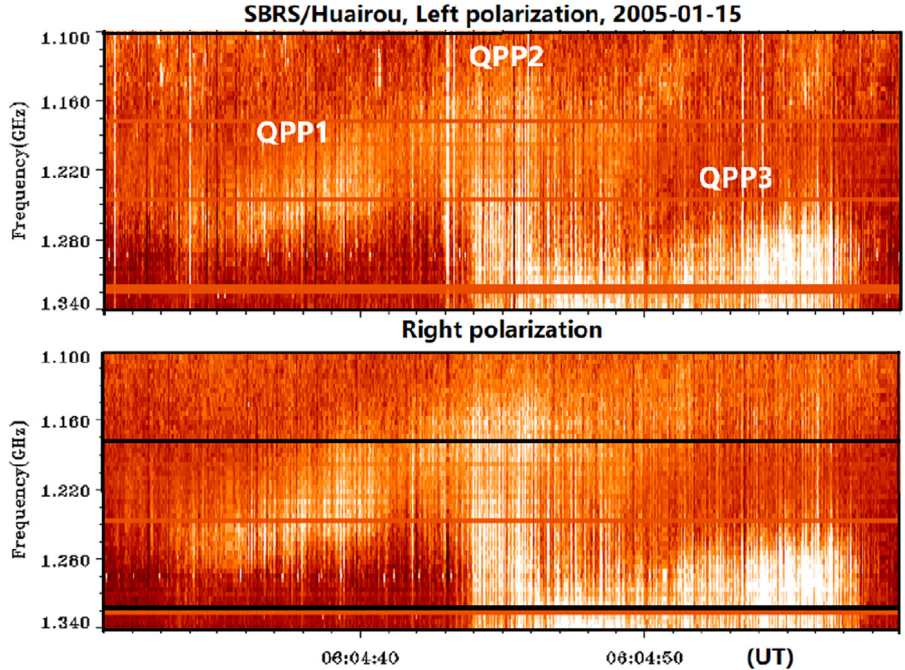

Fig. 5. The spectrogram of microwave quasi-periodic pulsation with zigzag pattern (Z-QPP) in the early rising phase of a M8.6 long-duration flare on 2005–1-15. The Z-QPP clearly consists of three sections with different global frequency drifting rates, bandwidths, and periods.

(3) The origin of microwave Zebra pattern bursts.

The microwave Zebra pattern (ZP) is the most complex, intriguing and very interesting spectral phenomenon in solar radio astronomy. It consists of several or tens of almost parallel and equidistant stripes on the broadband spectrogram and always related to the eruptive process of solar flares (Chernov, 2010; Tan et al., 2014; Zlotnik et al., 2014; Kaneda et al., 2017). The study of ZP structures took more than half a century, but the formation mechanism of ZP structures is still not clear. There are more than a dozen proposed theoretical models to explain the formation of ZP structures, including the Bernstein wave models (Rosenberg, 1972; Chiuderi et al., 1973; Zaitsev and Stepanov, 1983, etc.), whistler wave models (Kuijpers, 1975; Chernov et al., 2018), double plasma resonance models (DPR model, Winglee and Dulk, 1986; Zlotnik et al., 2003; Yasnov and Karlicky, 2004; Kuznetsov and Tsap, 2007; Li et al., 2021), but so far, there is no universally accepted model, the fundamental reason is that we still do not know the exact location and spatial structure of the ZP source region.

Chen et al. (2011) use the interferometric observations with simultaneous high spectral(~1 MHz) and high time (20 ms) resolution of the Frequency-Agile Solar Radiotelescope Subsystem Testbed (FST) and the Owens Valley Solar Array (OVSA) to study a microwave ZP burst during the decay phase of an X1.5 flare on 2006 December 14. They found that the apparent ZP source size is around 35″, the source centroid drifts 8–9″ irregularly in time and in the direction across the six stripes from high to low frequency at a fixed time. They concluded that the ZP burst is consistent with a double-plasma resonance model in which the radio emission occurs in resonance layers where the upper-hybrid frequency is harmonically related to the electron cyclotron frequency in a coronal magnetic loop.

By using the new spectral imaging observations with high resolutions, we may obtain more accurate location and spatial structure of the ZP source regions, which will further refine the relevant theoretical models, and finally reveal the origin of ZP bursts.

(4) Particle acceleration and propagation associated with small-scale microwave bursts.

Small-scale microwave bursts include microwave dot, spike, and narrow-band type III bursts. They have very short timescales (from several to tens ms), narrow bandwidth (around 1% of the central frequency), and very high brightness temperatures, that appeared in large groups at all stages of solar flares (Fleishman et al., 2003; Tan, 2013; Tang et al., 2021). They are believed to be related to some strong but very small scale non-thermal energy release processes and particle accelerations, possibly the elementary bursts associated with flaring fragmentation (Benz, 1985; Messmer and Benz, 2000). Chen et al. (2015) applied the Karl G. Jansky Very Large Array (VLA) to obtain high-cadence radio spectral imaging observations of a group of radio spike bursts in the frequency range of 1.0–2.0 GHz. They found that the source region of spike bursts located above the top of a flaring plasma loop where flaring termination shock occurred. They suggested that the flaring termination shock should be responsible, at least in part, for accelerating energetic electrons in the flare.

In fact, radio spike bursts can be observed at higher frequencies (Tan et al., 2019a). Where are their emission sources? How are the associated high-energy particles accelerated? What are the real mechanisms of the relevant emission and energetic particle acceleration? How to give the exact number of relevant high-energy particles? Such questions require the new spectral imaging observations. Furthermore, because spike bursts have extremely narrow bandwidth, we even need to apply a combined spectral-imaging observation mode (means a joint observation of spectral radioheliographs and dynamic spectrometers) to make sure we can recognize spike bursts.

(5) The formation of microwave fiber bursts.

Solar radio fiber bursts are one kind of peculiar spectral fine structures that frequently occur in great groups associated to solar flares (Aurass and Kliem, 1992; Karlicky et al., 2013; Wang et al., 2017; Wan et al., 2021, etc.). They have intermediate frequency drifting rates which is much faster than the type II radio burst but much slower than the type III bursts (Benz and Mann, 1998). At present, there are two types of formation models of solar radio fiber bursts: one is based on whistler waves (Kuijpers, 1975; Mann et al., 1987), and the other is based on Alfven waves (Bernold and Treumann, 1983; Treumann et al., 1990) or fast magnetoacoustic sausage-mode waves (Kuznetsov, 2006; Karlicky et al., 2013). In the above two types of theoretical models, the frequency drift rate of fiber bursts is related to the magnetic field in the coronal source, therefore both of them can be used to diagnose the coronal magnetic fields. However, the above two types of models have not been widely accepted at present, partly because they could not present a convincing explanation for the intermediate frequency drifting rates.

As most people believed that type II radio bursts is linked with the motion of fast CME-related shock waves with velocity from several hundreds to above thousand km s$^{-1}$ in the corona and the type III bursts are generated from the fast motion of high-energy electron beams with velocity from 0.3c to about 0.7c (here c is the light speed) in some open magnetic fields. According to general reasoning, the speed of the driver of fiber bursts should be between the above two types of bursts, that is to say, it should reach several to tens of thousands km s$^{-1}$. Then, what is the driver of fiber bursts for their intermediate frequency drifting rates? Although observations and theoretical research of solar radio fiber bursts have been studied for decades, their microphysical process, emission mechanism, and especially the physical links with the flaring process still remain unclear.

(6) The physical mechanism of microwave type III burst group.

In solar physics, we know that one of the main products of solar flares is nonthermal energetic particles, which may carry up to half of energy released in the flaring processes. It is well accepted that the solar radio type III bursts are the signature of energetic electron beams with energy of more than 10 keV propagating in the open magnetic field of the solar atmosphere and interplanetary space (Lin and Hudson, 1971; Dulk et al., 1984; Aschwanden et al., 1995; Chen et al., 2013; Reid and Ratcliffe, 2014; Zhang et al., 2018; Tan et al., 2019b, etc.). Chen et al. (2013) applied the new technique of radio dynamic imaging spectroscopy of radio type III bursts observed by the upgraded Karl G. Jansky Very Large Array (VLA) to trace the propagating trajectories of electron beam in the corona. They found that these electron beams emanate from an energy release site located in the low corona at a height below 15 Mm, and propagate along a bundle of discrete magnetic loops upward into the corona, see Fig. 6.

So far, most of the solar radio type III bursts are recorded at the frequency below 2 GHz, and this is consistent with the explanation that type III bursts are signature of energetic electron beams propagating in open magnetic field. However, there is growing observational evidence that radio type III bursts can also occur in the centimeter wavelength with frequency above 2.0 GHz, and they are always appear in groups (Ning et al., 2005, etc.). However, it is hard to conceive of an open magnetic field existing in the source regions of radio emission at frequencies above 2.0 GHz. Then, what is the formation processes of the microwave type III burst groups? What is the physical connection of type III bursts between at high frequencies and at low frequencies? And what is the difference between their corresponding particle accelerations? The answers to these questions depend on the new spectral imaging observations with high resolutions.(7) The formation of many bizarre spectral fine structures of solar radio bursts.

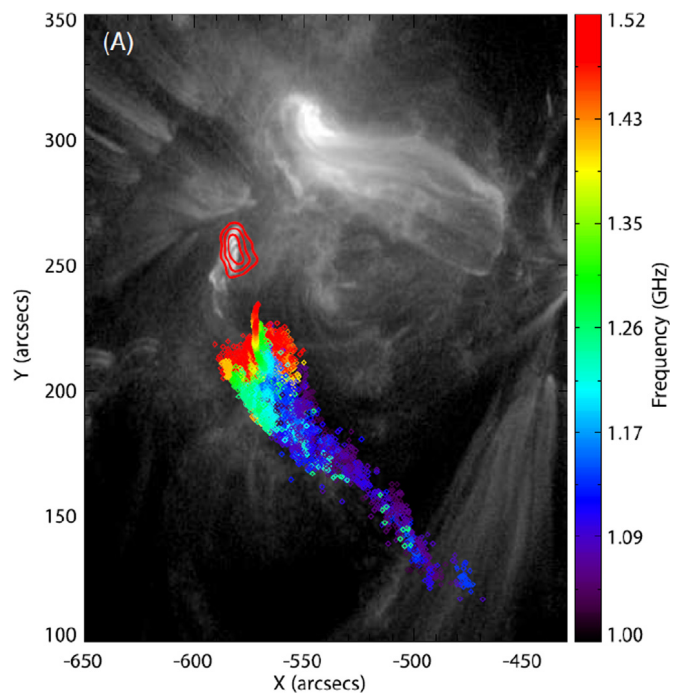

Fig. 6. The solar radio emission centroids of type III bursts observed by EVOSA, colored from blue to red in increasing frequencies showing electron beam trajectories in projection. The red contours are the 12 s integrated 12C25 keV HXR emission. Courtesy to Chen et al. (2013).

In radio observations of solar flares, it is frequently observed that there are some very strange and complex spectral fine structures which are short-lived strong emission and much different from the previous known structures (type I, II, III, IV and V bursts, ZP, fiber, QPP, spikes, etc.), some of them are complex with multiple ZPs, and some of them are completely bizarre with various complicated structures.

The left panels of Fig. 7 present a microwave burst with finger-shaped pattern occurred in the rising phase of an M1.1 flare on 2004 December 1, which is composed of several bright stripes with weak left polarization, including positive and negative drifting rates. It is possibly related to the magnetic reconnection and electron accelerations near the flaring source region. These energetic electrons propagated along different directions and generated different drifting rates, each direction implies a magnetized plasma loop.

The middle panels of Fig. 7 present a microwave burst with α-shaped pattern that occurred far from the impulsive phase of an X3.4 flare on 2006 December 13 with strong polarization, possibly related to one of small radio bursts. This pattern may be associated to fundamental plasma emission in some special magnetic loop configurations.

The right panels of Fig. 7 show a set of imbricate ZP bursts that occurred in the post-flare phase of an M3.5 flare on 2011 February 24. It is composed of three groups of similar ZP structures in imbrication, almost no polarization. According to DPR mechanism, this multiple ZP may be related to the trapped electrons in several adjacent plasma loops.

In addition to the problems we have mentioned above, there are still many strange and complex solar radio bursts whose formations are almost puzzled to us until now. Moreover, it is very likely that more novel solar radio bursts will be found in the future observations. What are the physical laws behind these phenomena? What do they mean for the generation and propagation of non-thermal particles? These questions also require the new spectral imaging observations. Revealing their formation mechanism and clarifying relevant physical laws will greatly promote the development of solar physics, astrophysics, and even plasma physics.

## 3. The possible application on space weather

The Sun is the ultimate power source in the solar system, and solar activity is the biggest disturbance source in the space environment. Solar flares, coronal mass ejections (CMEs) and solar energetic particle flows are the three main forms to trigger the disastrous space weather events. We know clearly that today radio observations are more used to monitor space weather than to predict space weather event. However, because radio observations are much more sensitive to the above three forms of solar activities than any other wavelength observations, they may play a crucial role in monitoring and predicting solar eruptions and therefore space weather hazards within few minutes to hours of time (Warmuth and Mann, 2005; Pick and Vilmer, 2005; Ndacyayisenga et al., 2021). Here, we will discuss the possible applications of radio observa-tions in cm - dm wavelength for space weather from three aspects: precursors of flares and CMEs, and the signatures of solar energetic particle flows. These discussions mainly focus on space weather science, and the results would be the input for the future space weather modeling and services.

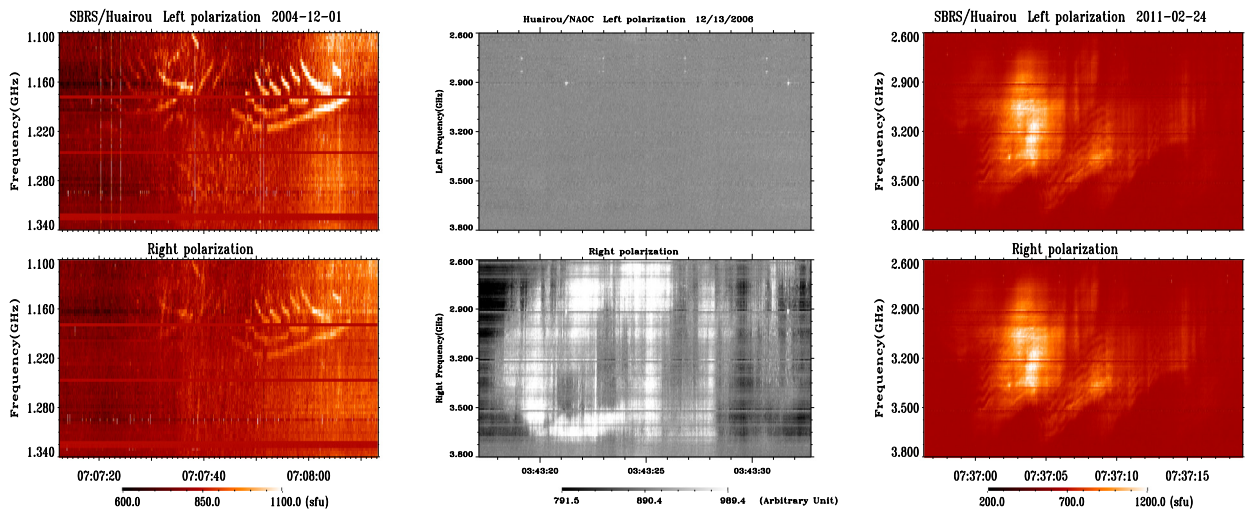

Fig. 7. Left panels show a microwave burst with finger-shaped pattern occurred in the rising phase of an M1.1 flare on 2004 December 1. The middle panels present a microwave burst with α-shaped pattern occurred far from the impulsive phase of an X3.2 flare on 2006 December 13. The right panels present a set of imbricate ZP bursts occurred in the post-flare phase of an M3.5 flare on 2011 February 24. All upper panels are the component of left polarization, and all lower panels are the component of right polarization.

*3.1. Radio precursors of solar flares*

As we know that solar flares are one of the most powerful initial driving sources of the disastrous space weather events. Any precursors of solar flares can be used to predict solar activities and the consequent disastrous space weather events (Joshi et al., 2011; Wang et al., 2017).

The precursors of solar flares may include the accumulation of magnetic energy, the changes of magnetic field strength and configuration, the motion of plasmas, and some small scale non-thermal energy release in the solar active regions. Radio emissions are sensitive to the above phenomena, especially at the dm-cm wavelengths. For example, the indices of solar radio flux at wavelength of 10.7 cm (2.8 GHz) ($F_{10.7}$) is widely used for ionospheric and upper-atmosphere modeling, also including the indices of $F_{11.1}$ (at 2.695 GHz) (Acebal and Sojka, 2011). They can use as a simple indicator of solar activity level for their high sensitivity (Tapping, 2013). In fact, both $F_{10.7}$ and $F_{11.1}$ are very sensitive to the variations of magnetic fields and non-thermal energy release in the upper chromosphere and low corona where source regions of most solar flares locate in. However, there are also many solar eruptions generating from higher or lower than the above regions, and their response to $F_{10.7}$ or $F_{11.1}$ index is relatively weak. The microwave type III pair bursts, which are explained as being produced by bi-directional electron beams generating from the site of magnetic reconnection and particle acceleration, are regarded as the start of solar eruptions and their separation frequency should be an indicator of the reconnection site (Aschwanden and Benz, 1997; Tan et al., 2016a, etc.). Recent statistical analysis of type III pairs shows that the separation frequency is in the range of about 1.08–3.42 GHz in different flare events (Tan et al., 2016a), while the previous statistics of Aschwanden and Benz (1997) showed that the separation frequency was in a relatively lower frequency range (from 220 MHz to 910 MHz). Because these two statistics are based on observations of type III bursts in different frequency ranges (3 GHz - 100 MHz on one hand and 7.6 GHz - 800 MHz on the other hand). Combined with these two statistical results, we believe that the separation frequency of radio type III pairs should be distributed in a much wide frequency range from less than several hundred MHz to more than 3 GHz in different flare events. Therefore, it is far from enough to use simply one index of $F_{10.7}$ or $F_{11.1}$ to predict the occurrence of solar flares, which will omit those eruptions that occur in the higher atmosphere with relatively low separation frequency or in the lower atmosphere with relatively high separation frequency.

A natural and reasonable selection is to use the combination of observation data at multiple frequencies to be the indicator for predicting the onset of solar eruptions. For example, the Next Generation Solar Flux Monitor (NGSFM, see Tapping and Morton, 2013) is being built and will measure multiple radio fluxes at 1.4, 1.6, 2.8, 4.9, 8.3, and 10.6 GHz. Such multiple frequency assembly may cover almost all source regions from the lower chromosphere to the high corona and provide comprehensive precursor information of solar flares. In practice, we can directly extract the radio flux at the above frequencies from the broadband dynamic spectrum observations. Furthermore, we can also extract the emission flux at any frequency. Through long-term continuous observation and statistical analysis, it is possible to find a further optimized frequency combination to be the indicator for predicting solar eruptions. In particular, by using broadband spectral-imaging observations (such as MUSER, EOVSA, and SRH, etc.), we can not only obtain the radio fluxes at required multiple frequencies, but also obtain information about the location of the flaring active regions. This is very important for obtaining comprehensive and accurate radio precursor of solar flares and for predicting their occurrence.

Besides the radio flux intensities, the spectral fine structures on the solar radio broadband dynamic spectrum may contain much more physical information of the flaring source region. Zhang et al. (2015) reported microwave QPP with period of 0.1–0.3 s and a dot burst group at the frequency range of 1.0–2.0 GHz that occurred in the preflare phases in several solar flares. These spectral fine structures have only relatively weak intensities and very short timescales. So far, we have found this kind of preflare spectral fine structures only in a few flare events, and it is not clear whether they are or not universal in other events. Usually, the single-dish solar radio telescope does not have the enough sensitivity to pick up such weak signals, while the solar radio antenna arrays will have much higher sensitivity which may pick up the above weak signals. With the new generation solar radioheliographs (MUSER, SRH, and the future FASR, et al.), it is possible to observe the radio precursors of solar flares directly from the active regions.

In addition, in recent years, many people have found that there are preflare QPP from the observations of soft X-ray (Tan et al., 2016b), Hα emission (Li et al., 2020) and other multiple wavelengths analysis. Are there corresponding radio counterparts of these preflare QPP? The dm-cm wavelength is the most promising frequency band to find such observation evidence, and it can also obtain more dynamic characteristics (e.g. the frequency drifting rate may provide the moving feature of the source region, the bandwidth may reflect its scale length, the polarization degree may contain the information of magnetic field, lifetimes, etc.) than other wavelengths. These dynamic characteristics are always related to the origin of preflare QPP and the physical links to the flare precursors, and therefore are meaningful for predicting the onset of solar flares.

From the above discussion, it is not difficult to find that the following parameters can be used to completely describe the characteristics of any solar radio bursts (Tan, 2008), including the radio precursors of solar flares.

a. Emission flux at multiple frequencies ($F$), including $F_{10.7}, F_{11.1}$ and the emission flux at other multiple frequencies mentioned above.

b. Frequency drifting rate ($\frac{df}{dt}$), including the frequency drifting rate of single burst and the global frequency drifting rate of a group of bursts.

c. Frequency bandwidth ($\triangle f$).

d. Polarization degree ($r$), usually is the degree of right-handed or left-handed circular polarization (RCP or LCP).

e. Timescale ($\tau$), including the lifetime of single bursts, period and duration of QPP, etc.

These parameters can be extracted directly from the dm-cm broadband spectral-imaging observations on the full solar disk, or active regions, or flaring source regions, and even around the coronal plasma loops. They can be regarded as the operational parameters for the prediction of solar flares. We know that these parameters are directly related to the precursor processes of solar flares, but so far, we do not know how to use these observational parameters to predict the occurrence of solar flare? This is still a difficult problem at present, which requires much more observation and more in-depth and detailed researches in the near future. Here, it is also necessary to develop new data analysis and new research methods.

### 3.2. Radio precursors of CMEs

The CME is another kind of powerful eruption that occurs in the solar corona, which may cast a huge cloud of plasma into interplanetary space with high speed. High-speed CME can drive shock waves in corona and interplanetary space and cause secondary acceleration of charged particles, and subsequently generate radio type II bursts in the low frequency range (from meter to kilometer wavelengths). When it reaches near-Earth space, it may trigger violent disturbances that may produce catastrophic space weather events (see the recent reviews of Vourlidas et al., 2020; Carley et al., 2020).

At present, predicting the occurrence of CME in advance is still a very complicated and difficult work. As we know that CMEs are driven by the explosive release of coronal magnetic energy. A major problem is that they are coronal phenomena but we have few direct measurements of the coronal magnetic field. As we discussed in Section 2, radio observations especially at dm- and cm-wavelengths may provide the most important diagnosing magnetic fields in and above the solar active regions, which may greatly improve the prediction level of CMEs. At the same time, because even mildly accelerated electrons of a few keV can also emit detectable radio emission, their detection could be an important precursor (Vourlidas et al., 2020). Huang et al. (2011) analyzed the radio emissions at different phases from the initiation to development of a CME and found that radio bursts have different spectral characteristics at different stages of the CME. Wang et al. (2011) statistically studied the relationship between CME and solar radio bursts, and found that most CME events are associated with radio type III bursts or radio spectral fine structures (FSs) during the initiation or early stages of the CMEs. This fact indicates that most CMEs contain the emissions of radio type III bursts/FSs before the onset of CMEs, independent of their fast or slow speeds. Aurass et al. (1999) proposed that the radio rising continuum should be another type of signature of CME. Sheiner and Fridman (2012) studied the sporadic radio bursts that precede geo-effective CMEs by using the observation data of solar broadband spectral observations in cm-, dm-, and meter-wave ranges during solar cycle 21–23, and revealed some regular features of sporadic microwave emissions observed before the geoeffective CMEs, which might be defined as radio precursors of CMEs.

But so far, the radio observation and the relevant research of the precursor of CMEs are very rare, and many problems remain to be clarified. For example, we are not sure what is the real radio precursor of CMEs especially in cm- and dm-wavelengths, and what is the physical mechanism of their formation? Radioheliographic observations at cm- and dm-wavelength will offer important information on the spatial and kinematic evolution of CME in its initiation and early phase of development. Naturally, here we also need to extract the operational parameters listed in the above paragraph from the observations. At the same time, we need to investigate the physical correlations between these parameters and the occurrence of CMEs. It is no doubt that more broadband spectral-imaging observations and more systematically statistical studies among the operational parameters and the onset of CMEs are crucial important for the prediction of space weather events.

### 3.3. Radio signature of solar energetic particle flows

The solar energetic particle flows are another important trigger of the disastrous space weather events which may produce serious impact on the electronic equipment of spacecrafts and even threaten astronauts outside the magnetosphere. Observations in X-rays, EUV and white light trace the plasma dynamics in the corona where particle acceleration takes place, while $\gamma$-ray, hard X-ray and radio observations can probe the acceleration of energetic particles in the source region and propagation in the corona and interplanetary space. Among them, radio emission provides key information related to the acceleration and propagation of energetic particles, such as type III bursts reveal the presence of magnetic field lines connecting the parent active region with the outer heliosphere, type II bursts should be an indicator of shock waves and fast CMEs, and type IV bursts reflect the confined electron populations in closed magnetic configurations (Klein et al., 2010; Kuroda et al., 2020; Ndacyayisenga et al., 2021).

Tracing back to the source, whether energetic particle flows, or the relevant radio type II, type III, or type IV bursts, they all originated from the primary energy release in the parent active regions. Therefore, the radio observation of the parent active region will give important information about the precursors, initial onset, and primary propagation characteristics of the solar energetic particle flows, which plays a key role in understanding the origin and impact of some disastrous space weather events. Among them, radio observations in the solar active regions are mainly concentrated in the cm- and dm-wavelengths. Here, it is necessary to mention the association between the radio type IV bursts and energetic particle flows. According to the early interpretations (Hakura and Goh, 1959; Wild et al., 1963), energetic electrons confined within a plasma cloud released from the flaring active region may produce type IV burst by synchrotron emission, and the turbulence within the plasma cloud will accelerate protons to high energies, which may escape from the plasma cloud when the energy was high enough. Such high energy protons may become the source of SEP events. The observations of radio type III pair bursts (Aschwanden and Benz, 1997; Tan et al., 2016a) show that frequency of radio bursts associated with the primary energy release are mainly located in cm- and dm-wavelengths. Additionally, many kinds of coherent radio spectral fine structures (such as spikes, dot, narrowband type III bursts, etc.) which also provide signatures of solar energetic particle flows accelerated by different processes (including magnetic reconnection acceleration, shock wave acceleration, and turbulence acceleration, etc.) in different conditions are also seen in dm - cm wavelengths. Additionally, the absence of radio type III burst should be an indicator of confined flare where the magnetic structure surrounding the region of energy release remains intact without observed CMEs (Wang and Zhang, 2007). Confined flares can be significant electron accelerators observed by the associated microwave or hard X-ray bursts, but without radio signatures of electrons escaping to the high corona (Klein et al., 2010).

With the new spectral imaging observations of the parent active regions at dm-cm wavelengths, we may not only obtain useful information about particle acceleration and propagation, but also provide the precursory characteristics of energetic particle flows from the source regions. It is also possible to get more accurate information about the solar energetic particle flows and predict their propagation and arrival to the near-Earth space.

## 4. Summary

Radio emissions are most sensitive to almost all nonthermal processes which occur in solar atmosphere. These nonthermal processes include various scales of violent energy release which take place in solar chromosphere to corona, such as solar flares, CMEs, eruptive filaments, plasma jets and particle accelerations, also include shock waves, and even various plasma instabilities. At the same time, radio emissions have sensitively response to the magnetic fields and their variations in the chromosphere and corona. We may apply solar radio observations to diagnose the magnetic fields in the corona, which is very difficult to obtain from the optical methods. The study of the above phenomena constitutes the main content of solar radio astronomy. Obviously, solar radio astronomy plays an irreplaceable role in the studies of solar physics, the modern astrophysics and even the possible applications on space weather.

With the advent of the new spectral imaging observations at dm-cm wavelengths with high spatial-spectral-temporal resolutions obtained from MUSER, EOVSA, SRH and even the future FASR, we may get considerable progress on many important problems on the frontier. These progress will include:

(1) We may diagnose coronal magnetic fields directly from the broadband radio spectral-imaging observations, and this may help us to reveal the energy accumulation and release in the source region of solar eruptions much more exactly. At the same time, the coronal magnetic field maps at different heights may help us to resolve the mystery of coronal heating.

(2) We will gain a new understanding of some basic problems of solar radio astronomy and solar physics, including the formation mechanism of many spectral fine structures of radio bursts, such as microwave broadband QPP, microwave Zebra pattern structures, fiber bursts, spike burst group, and other bizarre radio spectral fine structures. They will provide more exact details of the primary energy release, triggering mechanism, precursors, magnetic reconnections, and particle acceleration and propagations associated to solar flares and CMEs. Some problems also involve some basic problems in plasma physics, such as wave-particle interaction, wave-wave coupling, plasma heating and MHD turbulence, and so on.

(3) We will obtain more accurate understanding of the precursors of solar flare, CMEs and solar energetic particle flows, and this point is very important, because we may apply this knowledge to predict the disastrous space weather events, and provide service for the modern society with versatile high-tech systems.

From the broadband spectral-imaging observations at dm-cm wavelengths, we may obtain all operational parameters of solar radio bursts and radio precursors of solar flares, CMEs, and solar energetic particle flows, including emission flux, frequency bandwidth, frequency drifting rate, polarization degrees, and timescales. These parameters contain information about the origin, triggering mechanism, energy releasing and particle acceleration of solar eruptions, but so far we cannot give a clear and quantitative relationship between these parameters and solar eruptions. Therefore, at present, we still have no operational method to predict solar eruptions and the subsequent dis-astrous space weather events based on the above parameters. This requires much more observations and in-depth researches in the near future. Recently, some people introduced deep-learning algorithms to identify precursors of solar flare by using time series of SDO/HMI images and got a good results (Chen et al., 2019). This kind of method can also be applied to the study of radio parameters on the precursors of solar eruptions and space weather events.

**Declaration of Competing Interest**

The authors declare that they have no known competing financial interests or personal relationships that could have appeared to influence the work reported in this paper.

**Acknowledgments**

The authors would thank the referees greatly for their helpful and valuable comments to improve the manuscript of this paper. This work is supported by National Key R&D Program of China 2021YFA1600503, NSFC Grants 11973057, 12003048, 11790301 and 11941003, and the International Partnership Program of Chinese Academy of Sciences (183311KYSB20200003).